\documentclass[dvips]{acta}
\usepackage{supertabular,lscape,epsfig}
\usepackage{amssymb}
\usepackage{amsmath}
\DeclareSymbolFont{ppa}{OT1}{ppl}{m}{it}
\DeclareMathSymbol{\vv}{\mathalpha}{ppa}{'166}

\newfont{\hb}{rphvb at 10pt}
\newfont{\hbo}{rphvbo at 10pt}
\newfont{\bitt}{rptmbi at 12pt}
\newfont{\bits}{rptmbi at 11pt}

\SetPages{329}{343}

\SetVol{58}{2008}

\begin{document}

\newcommand{\TabCapp}[2]{\begin{center}\parbox[t]{#1}{\centerline{
  \small {\spaceskip 2pt plus 1pt minus 1pt T a b l e}
  \refstepcounter{table}\thetable}
  \vskip2mm
  \centerline{\footnotesize #2}}
  \vskip3mm
\end{center}}

\newcommand{\TTabCap}[3]{\begin{center}\parbox[t]{#1}{\centerline{
  \small {\spaceskip 2pt plus 1pt minus 1pt T a b l e}
  \refstepcounter{table}\thetable}
  \vskip2mm
  \centerline{\footnotesize #2}
  \centerline{\footnotesize #3}}
  \vskip1mm
\end{center}}

\newcommand{\MakeTableSepp}[4]{\begin{table}[p]\TabCapp{#2}{#3}
  \begin{center} \TableFont \begin{tabular}{#1} #4 
  \end{tabular}\end{center}\end{table}}

\newcommand{\MakeTableee}[4]{\begin{table}[htb]\TabCapp{#2}{#3}
  \begin{center} \TableFont \begin{tabular}{#1} #4
  \end{tabular}\end{center}\end{table}}

\newcommand{\MakeTablee}[5]{\begin{table}[htb]\TTabCap{#2}{#3}{#4}
  \begin{center} \TableFont \begin{tabular}{#1} #5 
  \end{tabular}\end{center}\end{table}}

\newfont{\bb}{ptmbi8t at 12pt}
\newfont{\bbb}{cmbxti10}
\newfont{\bbbb}{cmbxti10 at 9pt}
\newcommand{\uprule}{\rule{0pt}{2.5ex}}
\newcommand{\douprule}{\rule[-2ex]{0pt}{4.5ex}}
\newcommand{\dorule}{\rule[-2ex]{0pt}{2ex}}
\def\thefootnote{\fnsymbol{footnote}}
\begin{Titlepage}
\Title{The Optical Gravitational Lensing Experiment.\\
OGLE-III Photometric Maps of the Small Magellanic Cloud\footnote{Based
on observations obtained with the 1.3~m Warsaw telescope at the Las
Campanas Observatory of the Carnegie Institution of Washington.}}
\Author{A.~~U~d~a~l~s~k~i$^1$,~~I.~~S~o~s~z~y~\'n~s~k~i$^1$,
~~M.\,K.~~S~z~y~m~a~{\'n}~s~k~i$^1$,~~M.~~K~u~b~i~a~k$^1$, 
~~G.~~P~i~e~t~r~z~y~\'n~s~k~i$^{1,2}$,~~\L.~~W~y~r~z~y~k~o~w~s~k~i$^3$,
~~O.~~S~z~e~w~c~z~y~k$^2$, ~~K.~~U~l~a~c~z~y~k$^1$~~
and ~~R.~~P~o~l~e~s~k~i$^1$}
{$^1$Warsaw University Observatory, Al.~Ujazdowskie~4, 00-478~Warszawa, Poland\\
e-mail: (udalski,soszynsk,msz,mk,pietrzyn,kulaczyk,rpoleski)@astrouw.edu.pl\\
$^2$ Universidad de Concepci{\'o}n, Departamento de Fisica,
Casilla 160--C, Concepci{\'o}n, Chile\\
e-mail: szewczyk@astro-udec.cl\\
$^3$ Institute of Astronomy, University of Cambridge, Madingley Road,
Cambridge CB3~0HA,~UK\\
e-mail: wyrzykow@ast.cam.ac.uk}
\Received{December 27, 2008}
\end{Titlepage}

\Abstract{We present OGLE-III Photometric Maps of the Small Magellanic
Cloud. They contain precise, calibrated {\it VI} photometry of about 6.2
million stars from 41 OGLE-III fields in the SMC observed regularly in the
years 2001--2008 and covering about 14 square degrees in the sky. Also
precise astrometry of these objects is provided.
One of the fields, SMC140, is centered on the 47~Tucanae Galactic globular
cluster providing unique data on this object.

We discuss quality of the data and present a few color--magnitude
diagrams of the observed fields.

All photometric data are available to the astronomical community from the
OGLE Internet archive.}{Magellanic Clouds -- Surveys -- Catalogs --
Techniques: photometric}

\Section{Introduction}
One of the very important results of the second phase of the Optical
Gravitational Lensing Experiment (OGLE-II) -- long term, large scale
photometric survey -- was the publication of the OGLE Photometric Maps of
Dense Stellar Regions observed regularly in the course of the project. The
maps contain calibrated {\it BVI} or {\it VI} photometry and precise
astrometry of millions of stars from astrophysically so important objects
like the Large and Small Magellanic Clouds and Galactic Center
(Udalski \etal 1998, 2000, 2002).

OGLE maps are a very important source of astrophysical information and were
widely used by astronomers to many projects (\eg Evans \etal 2005,
Coe \etal 2005, Chiosi and Vallenari 2007, Haberl, Eger and Pietsch
2008). The maps constitute also a huge set of secondary photometric
standards and can be used for calibrating photometry.

The OGLE-II maps cover mostly the central parts of the Magellanic Clouds
and about 10 square degrees of the Galactic bulge. Large upgrade of the
OGLE project to OGLE-III phase in 2001 enabled regular coverage of much
larger area in the sky. After several years of regular monitoring of
OGLE-III targets the collected images were finally reprocessed to derive
the final photometry (Udalski \etal 2008a). This enabled construction of
new version of the maps based on OGLE-III observations. Recently the first
part of the new OGLE-III map set -- the maps of the Large Magellanic Cloud
-- has been released (Udalski \etal 2008b).

This paper is the next of this series. We present here new OGLE-III
photometric maps of the Small Magellanic Cloud. The new maps constitute a
significant extension to the OGLE-II maps of the SMC as they cover much
larger area and contain many more objects. Additionally they also include
the region of the very important Galactic globular cluster -- 47~Tuc.

The maps are available to the astrophysical community from the OGLE
Internet archive.

\Section{Observational Data}
Observational data presented in this paper were collected during the
OGLE-III phase between June 2001 and January 2008. 1.3-m Warsaw Telescope
at Las Campanas Observatory, Chile, operated by the Carnegie Institution of
Washington, equipped with the eight chip mosaic camera (Udalski 2003) was
used. One image covers approximately $35\times35$~arcmins on the sky with
the scale of 0.26~arcsec/pixel.

Observations were carried out in {\it V-} and {\it I}-band filters closely
resembling the standard bands. One should be however aware, that the OGLE
glass {\it I}-band filter approximates well the standard one for
$V-I<3$~mag colors. For very red objects the transformation to the standard
band is less precise. The vast majority of observations were obtained
through the {\it I}-band filter. Typically several hundred images for each
field were collected in this band and about 40--50 in the {\it V}-band. The
exposure time was 180 and 240 seconds for the {\it I-} and {\it V}-band,
respectively.

To obtain precise photometry in the dense regions of the SMC, observations
were conducted only in good seeing (less than 1\zdot\arcs8) and
transparency conditions. The median seeing of the {\it I}-band images of
the SMC collected during OGLE-III is equal to $1\zdot\arcs2$.

\MakeTableSepp{cccr}{12.5cm}{OGLE-III Fields in the SMC}
{\hline
\noalign{\vskip3pt}
Field & RA       &   DEC   & $N_{\rm Stars}$ \\
    & (2000)   &  (2000) &  \\
\noalign{\vskip3pt}
\hline
\noalign{\vskip3pt}
SMC100 & 0\uph50\upm06\zdot\ups4 & $-73\arcd08\arcm19\arcs$ &  512714 \\
SMC101 & 0\uph50\upm04\zdot\ups0 & $-72\arcd32\arcm57\arcs$ &  342621 \\
SMC102 & 0\uph50\upm09\zdot\ups0 & $-71\arcd57\arcm09\arcs$ &  121100 \\
SMC103 & 0\uph50\upm08\zdot\ups6 & $-73\arcd43\arcm44\arcs$ &  274952 \\
SMC104 & 0\uph50\upm06\zdot\ups7 & $-74\arcd19\arcm21\arcs$ &  128812 \\
SMC105 & 0\uph57\upm50\zdot\ups8 & $-72\arcd44\arcm37\arcs$ &  365672 \\
SMC106 & 0\uph58\upm06\zdot\ups7 & $-73\arcd20\arcm21\arcs$ &  362026 \\
SMC107 & 0\uph58\upm23\zdot\ups0 & $-73\arcd55\arcm50\arcs$ &  142474 \\
SMC108 & 0\uph57\upm31\zdot\ups5 & $-72\arcd09\arcm29\arcs$ &  413846 \\
SMC109 & 0\uph57\upm18\zdot\ups8 & $-71\arcd33\arcm56\arcs$ &  132507 \\
SMC110 & 1\uph05\upm40\zdot\ups0 & $-72\arcd44\arcm35\arcs$ &  234190 \\
SMC111 & 1\uph06\upm10\zdot\ups2 & $-73\arcd20\arcm12\arcs$ &  181265 \\
SMC112 & 1\uph06\upm37\zdot\ups3 & $-73\arcd55\arcm58\arcs$ &  112770 \\
SMC113 & 1\uph05\upm02\zdot\ups8 & $-72\arcd09\arcm32\arcs$ &  281492 \\
SMC114 & 1\uph04\upm31\zdot\ups5 & $-71\arcd34\arcm06\arcs$ &  156581 \\
SMC115 & 1\uph13\upm22\zdot\ups4 & $-72\arcd44\arcm27\arcs$ &  110029 \\
SMC116 & 1\uph13\upm57\zdot\ups6 & $-73\arcd20\arcm36\arcs$ &  117909 \\
SMC117 & 1\uph15\upm03\zdot\ups4 & $-73\arcd55\arcm34\arcs$ &  47030  \\
SMC118 & 1\uph12\upm38\zdot\ups7 & $-72\arcd08\arcm56\arcs$ &  91419  \\
SMC119 & 1\uph11\upm47\zdot\ups8 & $-71\arcd34\arcm05\arcs$ &  102201 \\
SMC120 & 1\uph20\upm58\zdot\ups8 & $-72\arcd45\arcm10\arcs$ &  59056  \\
SMC121 & 1\uph22\upm04\zdot\ups2 & $-73\arcd20\arcm24\arcs$ &  31525  \\
SMC122 & 1\uph23\upm15\zdot\ups8 & $-73\arcd55\arcm54\arcs$ &  21627  \\
SMC123 & 1\uph20\upm05\zdot\ups9 & $-72\arcd09\arcm17\arcs$ &  34723  \\
SMC124 & 1\uph19\upm03\zdot\ups5 & $-71\arcd34\arcm06\arcs$ &  43827  \\
SMC125 & 0\uph42\upm05\zdot\ups2 & $-73\arcd19\arcm43\arcs$ &  384551 \\
SMC126 & 0\uph42\upm25\zdot\ups5 & $-72\arcd43\arcm46\arcs$ &  207374 \\
SMC127 & 0\uph42\upm58\zdot\ups5 & $-71\arcd08\arcm45\arcs$ &  21564  \\
SMC128 & 0\uph41\upm48\zdot\ups1 & $-73\arcd55\arcm17\arcs$ &  190156 \\
SMC129 & 0\uph41\upm35\zdot\ups4 & $-74\arcd30\arcm26\arcs$ &  56668  \\
SMC130 & 0\uph34\upm05\zdot\ups6 & $-73\arcd19\arcm15\arcs$ &  125840 \\
SMC131 & 0\uph34\upm33\zdot\ups6 & $-72\arcd43\arcm47\arcs$ &  71873  \\
SMC132 & 0\uph35\upm49\zdot\ups4 & $-71\arcd08\arcm24\arcs$ &  18600  \\
SMC133 & 0\uph33\upm36\zdot\ups8 & $-73\arcd55\arcm03\arcs$ &  101530 \\
SMC134 & 0\uph32\upm52\zdot\ups3 & $-74\arcd30\arcm24\arcs$ &  42505  \\
SMC135 & 0\uph26\upm02\zdot\ups8 & $-73\arcd19\arcm22\arcs$ &  54487  \\
SMC136 & 0\uph26\upm48\zdot\ups0 & $-72\arcd43\arcm52\arcs$ &  41378  \\
SMC137 & 0\uph28\upm41\zdot\ups8 & $-71\arcd08\arcm22\arcs$ &  15786  \\
SMC138 & 0\uph25\upm21\zdot\ups4 & $-73\arcd55\arcm19\arcs$ &  77662  \\
SMC139 & 0\uph24\upm16\zdot\ups3 & $-74\arcd30\arcm25\arcs$ &  33851  \\
SMC140 & 0\uph24\upm08\zdot\ups9 & $-72\arcd04\arcm54\arcs$ &  343389 \\
\noalign{\vskip3pt}
\hline}
\begin{landscape}
\begin{figure}[p]
\vskip12cm
\FigCap{OGLE-III fields in the SMC (black squares: 100--140) overplotted
on the image obtained by the ASAS all sky survey. Gray strips (1--11) mark\\
OGLE-II~fields.}
\end{figure}
\end{landscape}

Table~1 lists all the SMC fields observed during OGLE-III phase as well as
the equatorial coordinates of their centers and number of stars detected in
the {\it I}-band. Fig.~1 presents an image of the SMC folded from images
taken by the ASAS survey program (Pojmañski 1997) with contours of the
OGLE-III fields. The area of the sky covered by OGLE-III observations
reaches 14 square degrees.
\vspace*{9pt}
\Section{Photometric Maps of the SMC}
\vspace*{5pt}
Detailed description of the construction of the OGLE-III maps was presented
by Udalski \etal (2008b). The procedure for the SMC maps was identical as
in the case of the LMC. We only recall here that the maps contain the mean
photometry of all detected stellar objects.  The mean photometry was
obtained for all objects with minimum 4 and 6 observations in the {\it V-}
and {\it I}-band, respectively by averaging all observations after removing
$5\sigma$ deviating points.

Table~2 lists as an example the first 25 entries from the map of the
SMC100.1 subfield. The columns contain: (1)~ID number; (2,3)~equatorial
coordinates J2000.0; (4,5)~$X,Y$ pixel coordinates in the {\it I}-band
reference image; (6,7,8)~photometry: {\it V}, $V-I$, {\it I};
(9,10,11)~number of points for average magnitude, number of $5\sigma$
removed points, $\sigma$ of magnitude for {\it V}-band; (12,13,14)~same as
(9,10,11) for the {\it I}-band. 9.999 or 99.999 markers mean ``no
data''. $-1$ in column (9) indicates multiple {\it V}-band
cross-identification (the average magnitude is the mean of the merged
datasets).

The full set of the OGLE-III Photometric Maps of the SMC is available
from the OGLE Internet archive (see Section~5).

\vspace*{9pt}
\Section{Discussion}
\vspace*{5pt}
The new OGLE-III Photometric Maps of the SMC contain data for about 6.2
million stars located in 41 OGLE-III fields. The maps are significant
extension to the OGLE-II maps of this galaxy (Udalski \etal 1998) that
covered mostly the central bar. Therefore they are much better tool for
studying properties of the SMC.

Typical accuracy of the OGLE-III Photometric Maps is presented in Figs.~2
and 3 where standard deviation of magnitudes as a function of magnitude in
the {\it V-} and {\it I}-band is plotted. The plots are shown for the most
dense central subfield SMC100.2 and empty subfield SMC138.2 located at the
outskirts of this galaxy.

Completeness of the photometry can be assessed from the histograms
presented in Figs.~4 and 5 for the same two subfields. It reaches
$I\approx21$~mag and $V\approx21.5$~mag and, as expected, the photometry is
deeper and more complete for less crowded fields.

\begin{landscape}
\renewcommand{\arraystretch}{.95}
\MakeTableSepp{
r@{\hspace{17pt}}
c@{\hspace{17pt}}
c@{\hspace{17pt}}
r@{\hspace{17pt}}
c@{\hspace{17pt}}
c@{\hspace{17pt}}
r@{\hspace{17pt}}
c@{\hspace{17pt}}
c@{\hspace{17pt}}
r@{\hspace{17pt}}
c@{\hspace{17pt}}
r@{\hspace{17pt}}
c@{\hspace{17pt}}
c}{12.5cm}
{OGLE-III Photometric Map of the SMC100.2 subfield.}
{\hline
\noalign{\vskip4pt}
ID & RA    & DEC   & $X$~~~~ & $Y$ & $V$ & $V-I$ & $I$ & $N_V$ & $N^{\rm bad}_V$ & $\sigma_V$ & $N_I$ & $N^{\rm bad}_I$ & $\sigma_I$\\
   &(2000) & (2000)&&&&&&&&&&&\\
\noalign{\vskip4pt}
\hline
\noalign{\vskip4pt}
  1 & 0\uph50\upm06\zdot\ups29 & $-73\arcd16\arcm31\zdot\arcs4$ &  250.36 &~~18.29 & 13.693 & $-0.095$ & 13.788 & 15 & 0 & 0.012 & 432 & 0 & 0.012\\
  2 & 0\uph50\upm08\zdot\ups58 & $-73\arcd15\arcm53\zdot\arcs1$ &  397.52 &~~56.70 & 13.795 & $ 0.017$ & 13.778 & 38 & 0 & 0.011 & 613 & 0 & 0.010\\
  3 & 0\uph50\upm09\zdot\ups23 & $-73\arcd15\arcm15\zdot\arcs3$ &  542.31 &~~67.96 & 15.444 & $ 1.170$ & 14.274 & 42 & 0 & 0.007 & 620 & 0 & 0.007\\
  4 & 0\uph50\upm13\zdot\ups06 & $-73\arcd17\arcm10\zdot\arcs2$ &  101.28 & 130.12 & 13.707 & $ 0.791$ & 12.916 & 42 & 0 & 0.006 & 629 & 3 & 0.013\\
  5 & 0\uph50\upm23\zdot\ups93 & $-73\arcd13\arcm20\zdot\arcs6$ &  981.38 & 313.77 & 15.192 & $ 1.692$ & 13.500 & 43 & 0 & 0.014 & 632 & 0 & 0.010\\
  6 & 0\uph50\upm29\zdot\ups46 & $-73\arcd16\arcm39\zdot\arcs7$ &  217.32 & 402.07 & 16.467 & $ 2.289$ & 14.178 & 43 & 0 & 0.121 & 632 & 0 & 0.076\\
  7 & 0\uph50\upm34\zdot\ups15 & $-73\arcd13\arcm54\zdot\arcs2$ &  851.53 & 482.90 & 15.636 & $ 2.378$ & 13.258 & 43 & 0 & 0.241 & 632 & 0 & 0.072\\
  8 & 0\uph50\upm34\zdot\ups42 & $-73\arcd13\arcm00\zdot\arcs9$ & 1056.00 & 488.33 & 14.566 & $ 9.999$ & 99.999 & 43 & 0 & 0.010 &   0 & 0 & 9.999\\
  9 & 0\uph50\upm34\zdot\ups61 & $-73\arcd15\arcm52\zdot\arcs6$ &  397.47 & 488.27 & 15.038 & $ 1.046$ & 13.992 & 43 & 0 & 0.005 & 632 & 0 & 0.007\\
 10 & 0\uph50\upm37\zdot\ups20 & $-73\arcd17\arcm24\zdot\arcs0$ & ~~46.56 & 529.47 & 13.757 & $-0.081$ & 13.837 & 37 & 0 & 0.008 & 567 & 0 & 0.008\\
 11 & 0\uph50\upm36\zdot\ups96 & $-73\arcd13\arcm06\zdot\arcs3$ & 1035.11 & 530.49 & 15.252 & $ 1.345$ & 13.906 & 43 & 0 & 0.007 & 632 & 0 & 0.007\\
 12 & 0\uph50\upm38\zdot\ups03 & $-73\arcd16\arcm14\zdot\arcs7$ &  312.41 & 544.55 & 15.536 & $ 1.539$ & 13.997 & 43 & 0 & 0.011 & 632 & 0 & 0.008\\
 13 & 0\uph50\upm41\zdot\ups90 & $-73\arcd16\arcm50\zdot\arcs1$ &  176.32 & 608.02 & 14.996 & $ 1.557$ & 13.439 & 43 & 0 & 0.015 & 632 & 0 & 0.010\\
 14 & 0\uph50\upm43\zdot\ups63 & $-73\arcd16\arcm32\zdot\arcs0$ &  245.73 & 636.94 & 14.979 & $ 1.174$ & 13.804 & 43 & 0 & 0.006 & 632 & 0 & 0.006\\
 15 & 0\uph50\upm43\zdot\ups44 & $-73\arcd12\arcm55\zdot\arcs2$ & 1077.03 & 638.38 & 15.158 & $ 2.134$ & 13.023 & 43 & 0 & 0.069 & 632 & 0 & 0.031\\
 16 & 0\uph50\upm46\zdot\ups40 & $-73\arcd13\arcm15\zdot\arcs4$ &  999.46 & 687.22 & 14.784 & $ 1.556$ & 13.228 & 43 & 0 & 0.007 & 631 & 1 & 0.009\\
 17 & 0\uph50\upm53\zdot\ups29 & $-73\arcd16\arcm51\zdot\arcs4$ &  170.32 & 796.58 & 15.490 & $ 1.340$ & 14.150 & 43 & 0 & 0.006 & 632 & 0 & 0.006\\
 18 & 0\uph50\upm54\zdot\ups97 & $-73\arcd17\arcm04\zdot\arcs0$ &  121.81 & 824.01 & 13.975 & $ 9.999$ & 99.999 & 43 & 0 & 0.004 &   0 & 0 & 9.999\\
 19 & 0\uph50\upm55\zdot\ups61 & $-73\arcd12\arcm54\zdot\arcs9$ & 1076.78 & 840.67 & 14.259 & $ 0.553$ & 13.706 & 43 & 0 & 0.009 & 631 & 1 & 0.007\\
 20 & 0\uph50\upm56\zdot\ups57 & $-73\arcd12\arcm55\zdot\arcs5$ & 1074.43 & 856.59 & 14.800 & $ 0.668$ & 14.132 & 43 & 0 & 0.008 & 632 & 0 & 0.007\\
 21 & 0\uph50\upm57\zdot\ups41 & $-73\arcd17\arcm33\zdot\arcs4$ & ~~~8.61 & 863.70 & 16.565 & $ 2.900$ & 13.665 & 18 & 0 & 0.342 & 361 & 0 & 0.259\\
 22 & 0\uph50\upm58\zdot\ups01 & $-73\arcd17\arcm08\zdot\arcs6$ &  103.76 & 874.24 & 14.471 & $ 1.576$ & 12.894 & 42 & 0 & 0.007 & 629 & 3 & 0.012\\
 23 & 0\uph51\upm01\zdot\ups95 & $-73\arcd16\arcm06\zdot\arcs9$ &  340.06 & 941.04 & 14.394 & $ 9.999$ & 99.999 & 43 & 0 & 0.011 &   0 & 0 & 9.999\\
 24 & 0\uph50\upm05\zdot\ups91 & $-73\arcd14\arcm06\zdot\arcs5$ &  806.60 &~~13.74 & 16.592 & $ 1.483$ & 15.109 & 13 & 0 & 0.016 & 404 & 0 & 0.013\\
 25 & 0\uph50\upm06\zdot\ups38 & $-73\arcd13\arcm22\zdot\arcs3$ &  976.00 &~~22.13 & 16.515 & $ 1.331$ & 15.184 & 18 & 0 & 0.010 & 460 & 0 & 0.012\\
\noalign{\vskip4pt}
\hline}
\end{landscape}

\pagebreak
\begin{figure}[h]
\centerline{\includegraphics[width=8.7cm, bb=30 50 510 550]{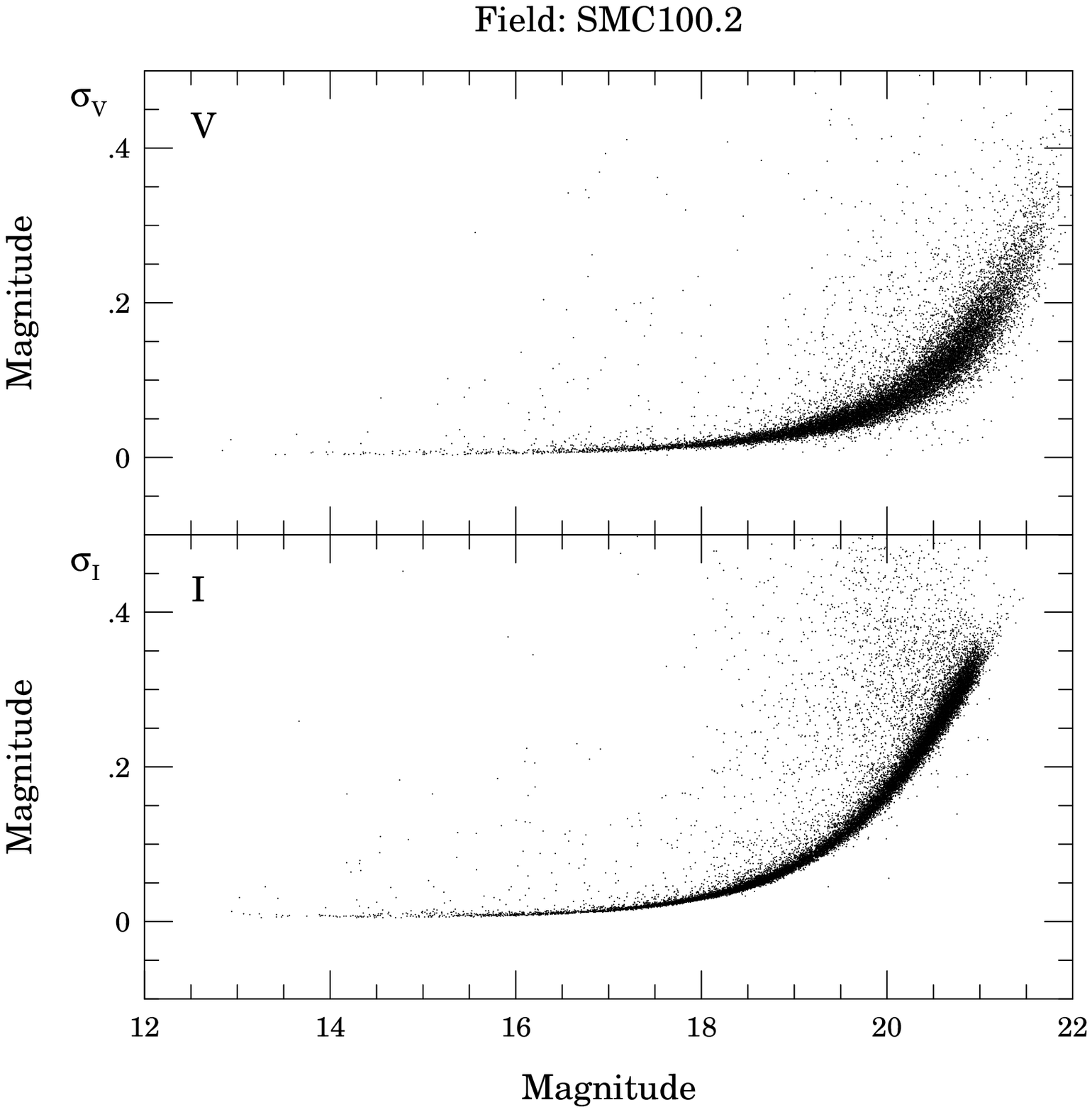}}
\FigCap{Standard deviation of magnitudes as a function of magnitude for
the central bar subfield SMC100.2.}
\centerline{\includegraphics[width=8.7cm, bb=30 50 510 550]{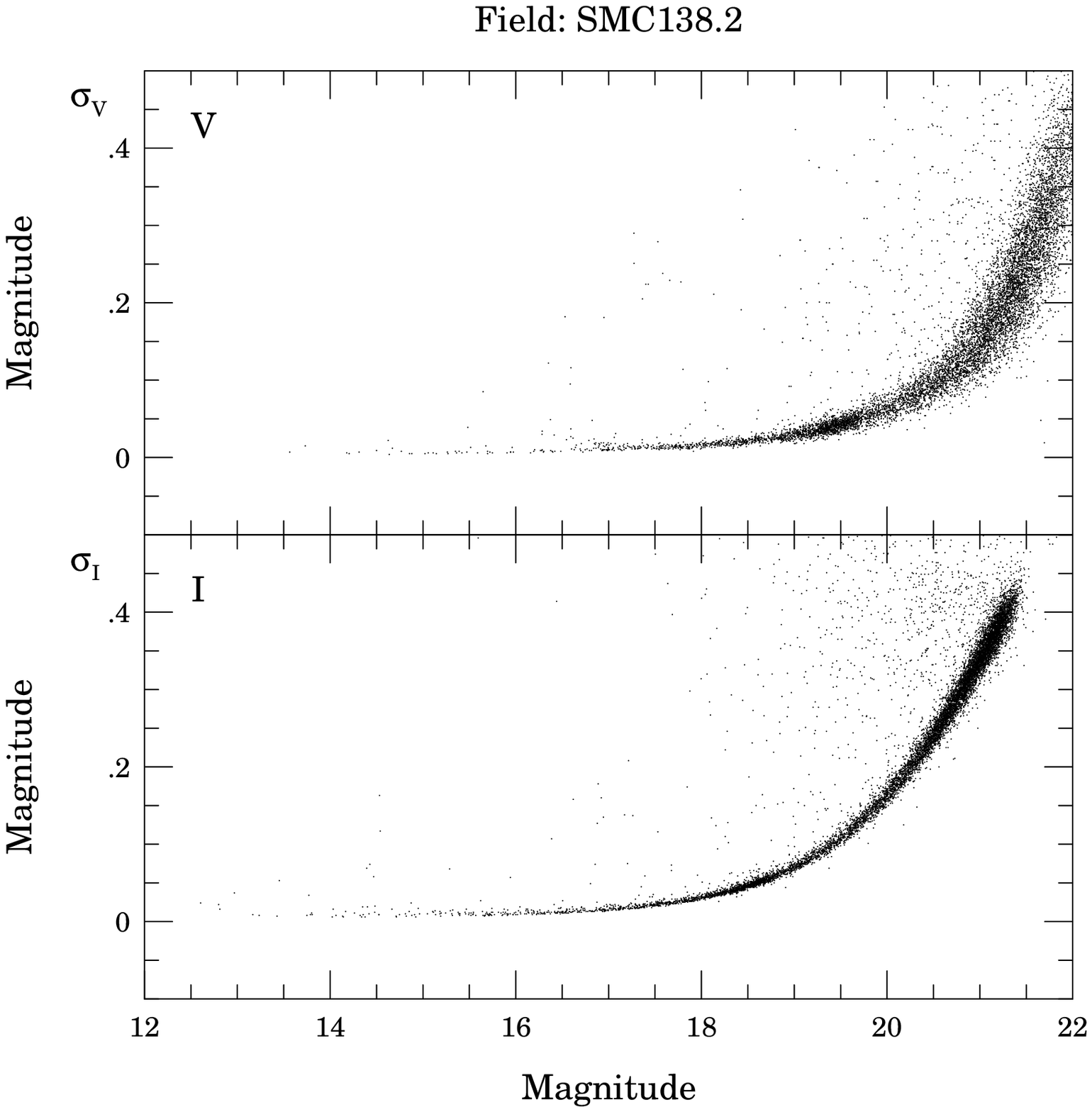}}
\FigCap{Same as in Fig.~2 for the subfield SMC138.2 located in the
outskirts of the SMC.}
\end{figure}
\begin{figure}[p]
\centerline{\includegraphics[width=9cm, bb=30 50 510 550]{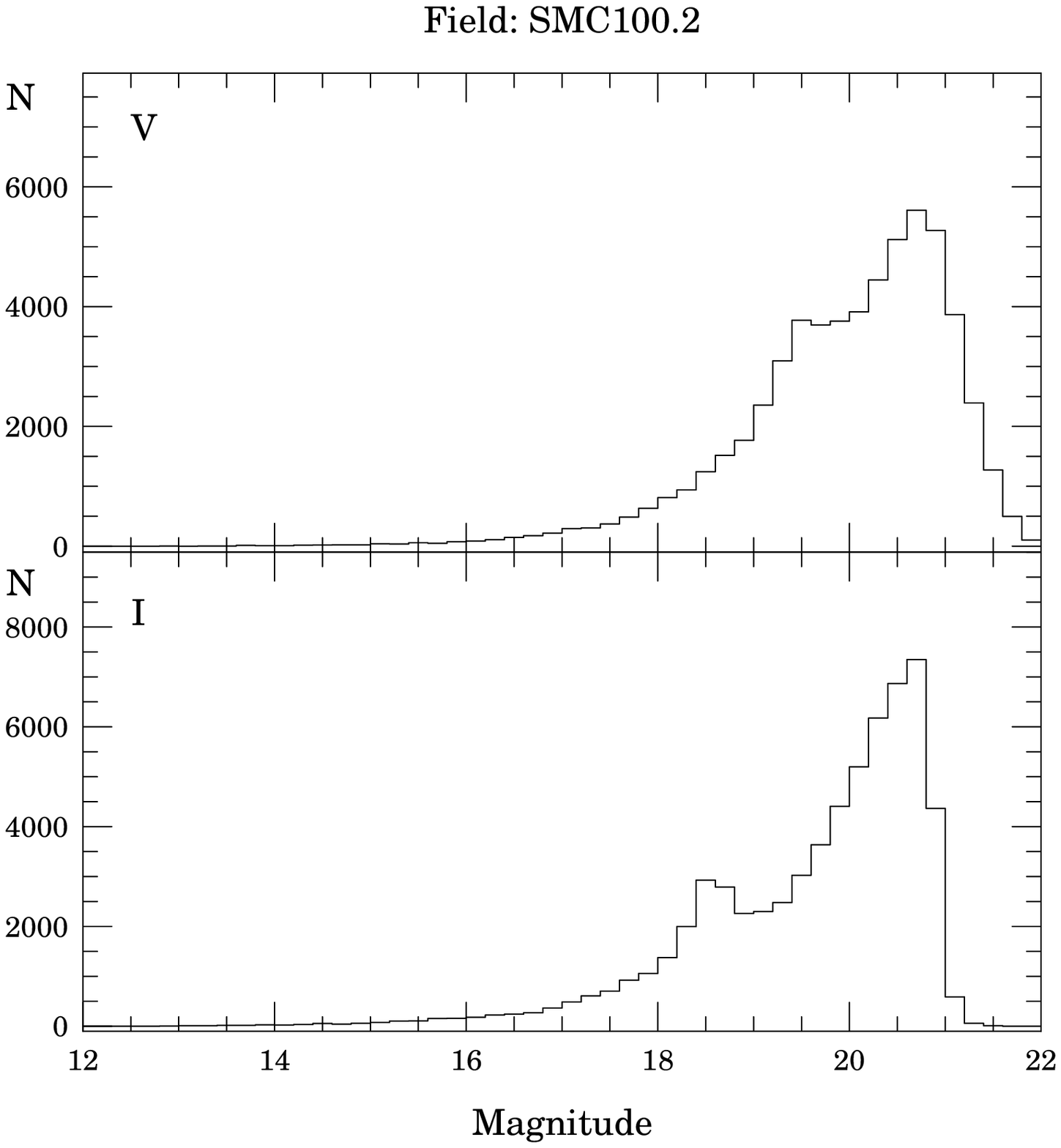}}
\FigCap{Histogram of magnitudes for the central bar subfield SMC100.2.}
\centerline{\includegraphics[width=9cm, bb=30 50 510 550]{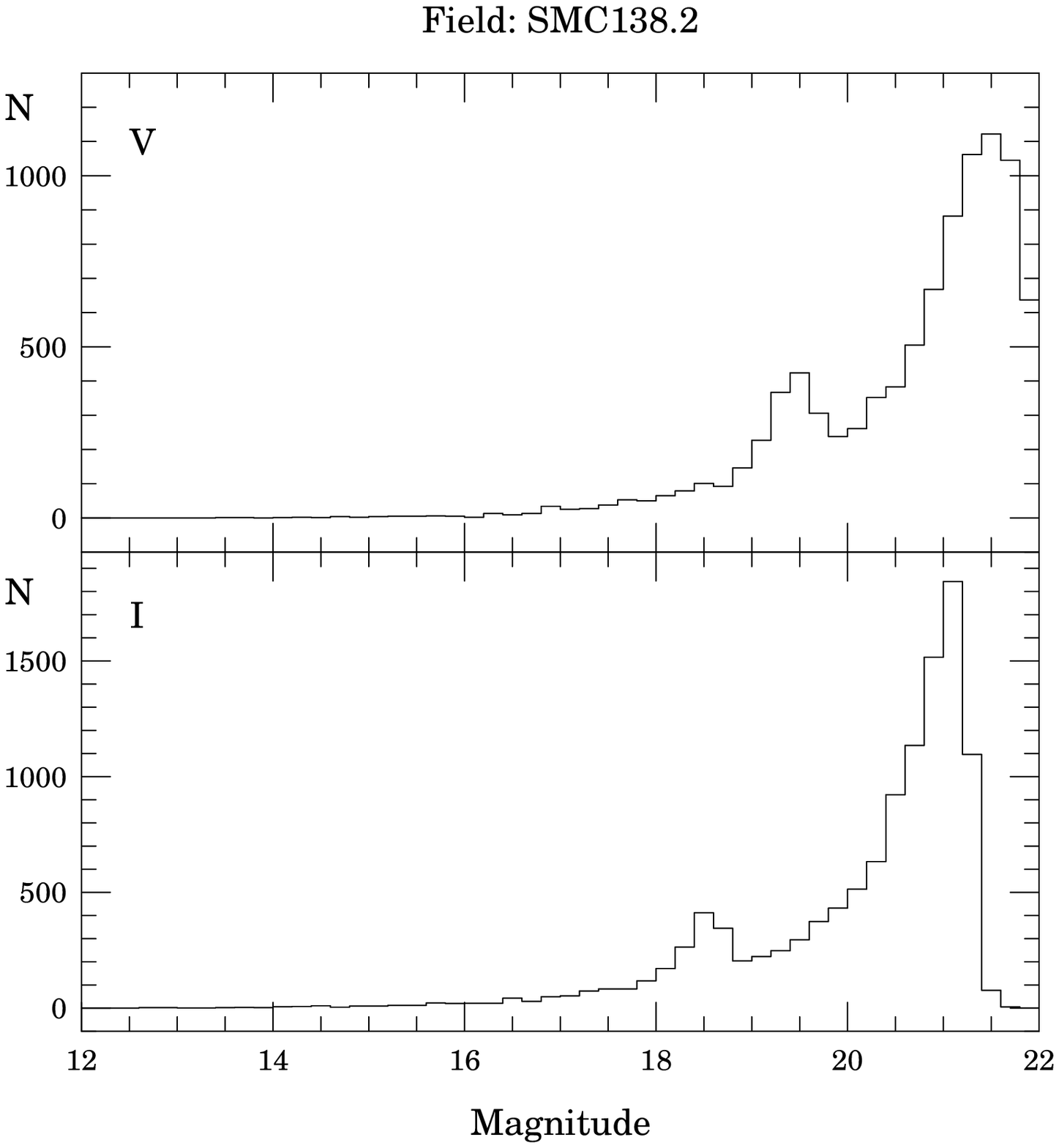}}
\FigCap{Same as in Fig.~5 for the outer subfield SMC138.2.} 
\end{figure}
\begin{figure}[p]
\centerline{\includegraphics[width=14.5cm, bb=10 50 560 750]{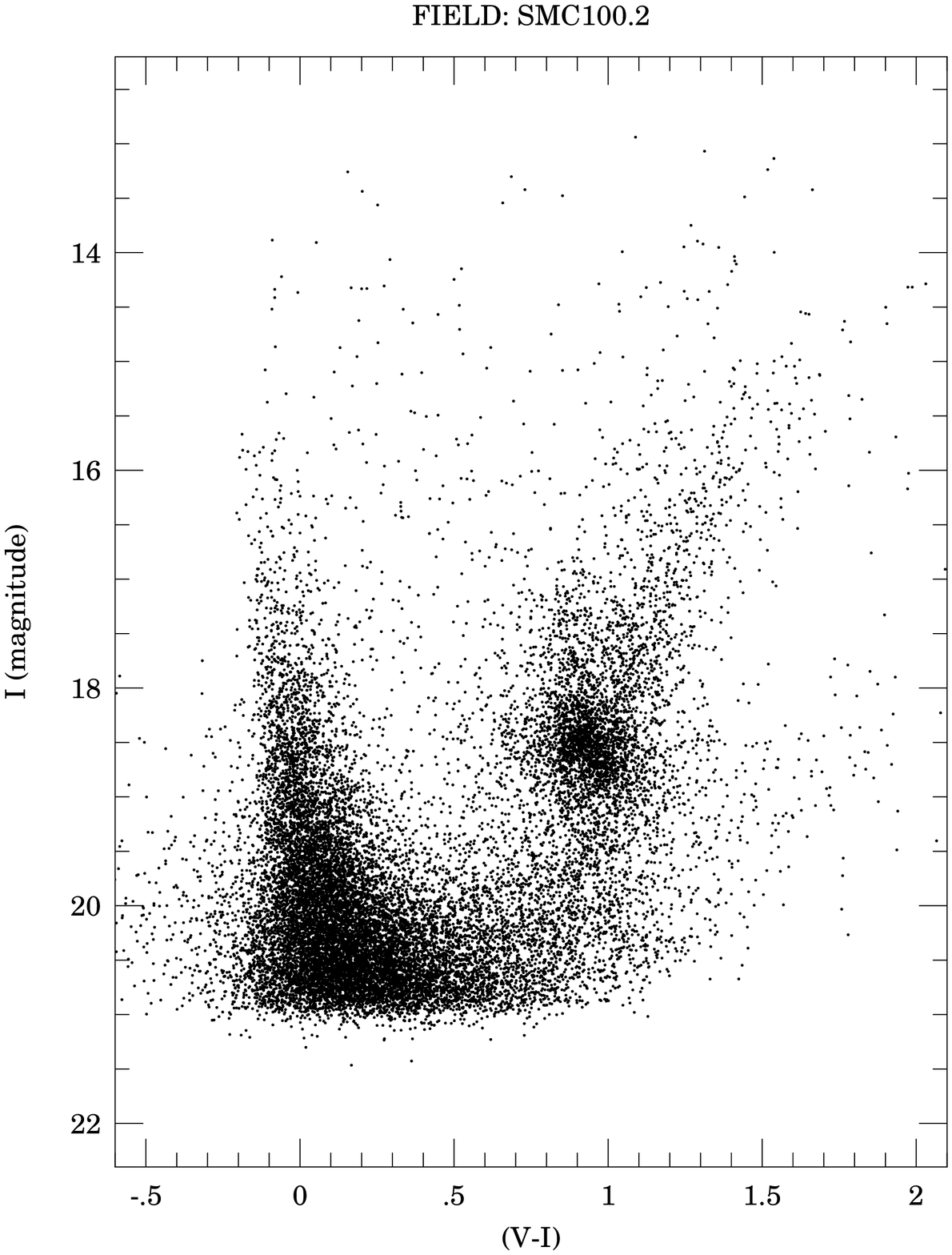}}
\FigCap{Color--magnitude diagram of the central bar subfield SMC100.2.
Only 30\% of stars are plotted for clarity.}
\end{figure}
\begin{figure}[p]
\centerline{\includegraphics[width=14.5cm, bb=10 50 560 750]{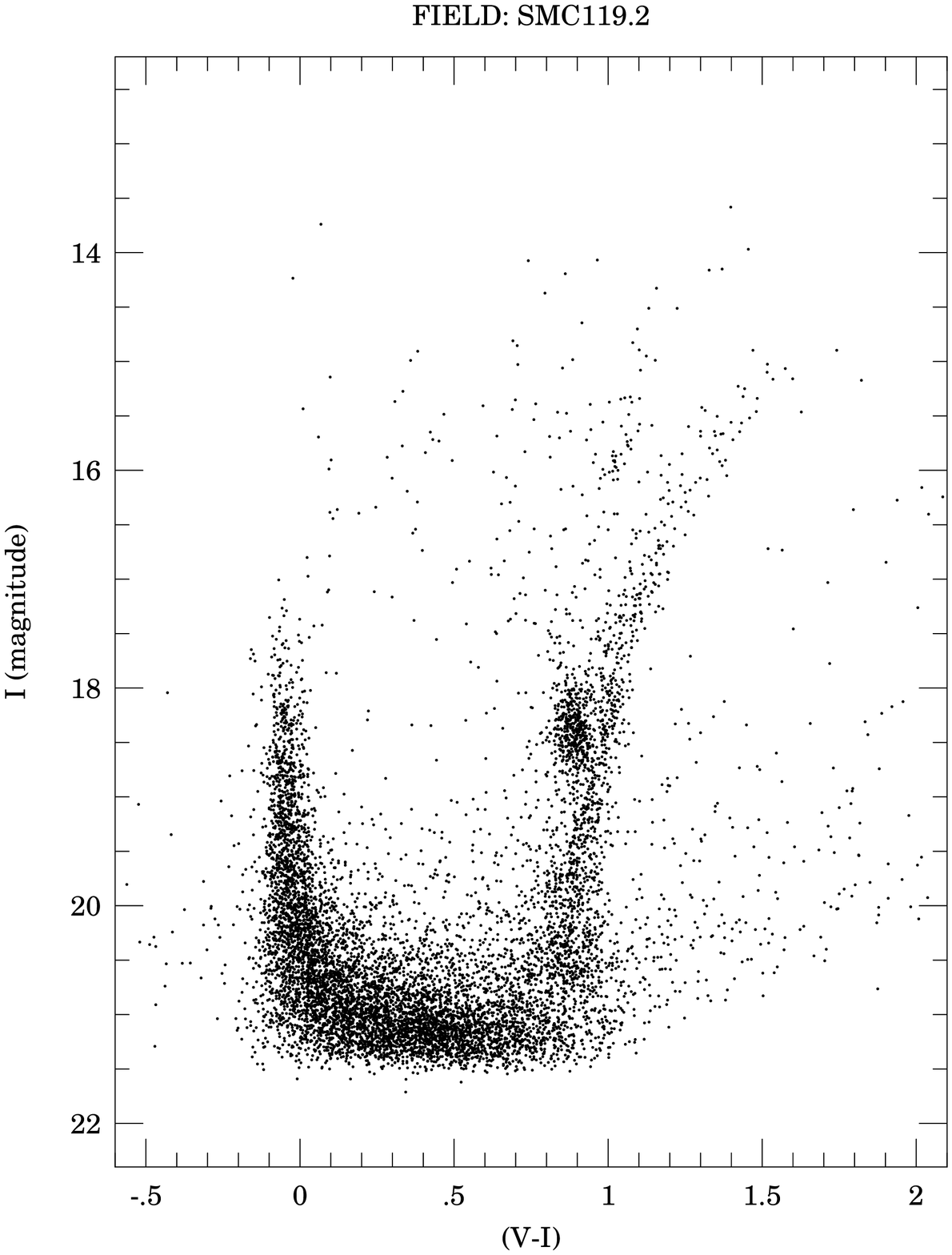}}
\FigCap{Color--magnitude diagram of the eastern wing subfield SMC119.2.}
\end{figure}
\begin{figure}[p]
\centerline{\includegraphics[width=14.5cm, bb=10 50 560 750]{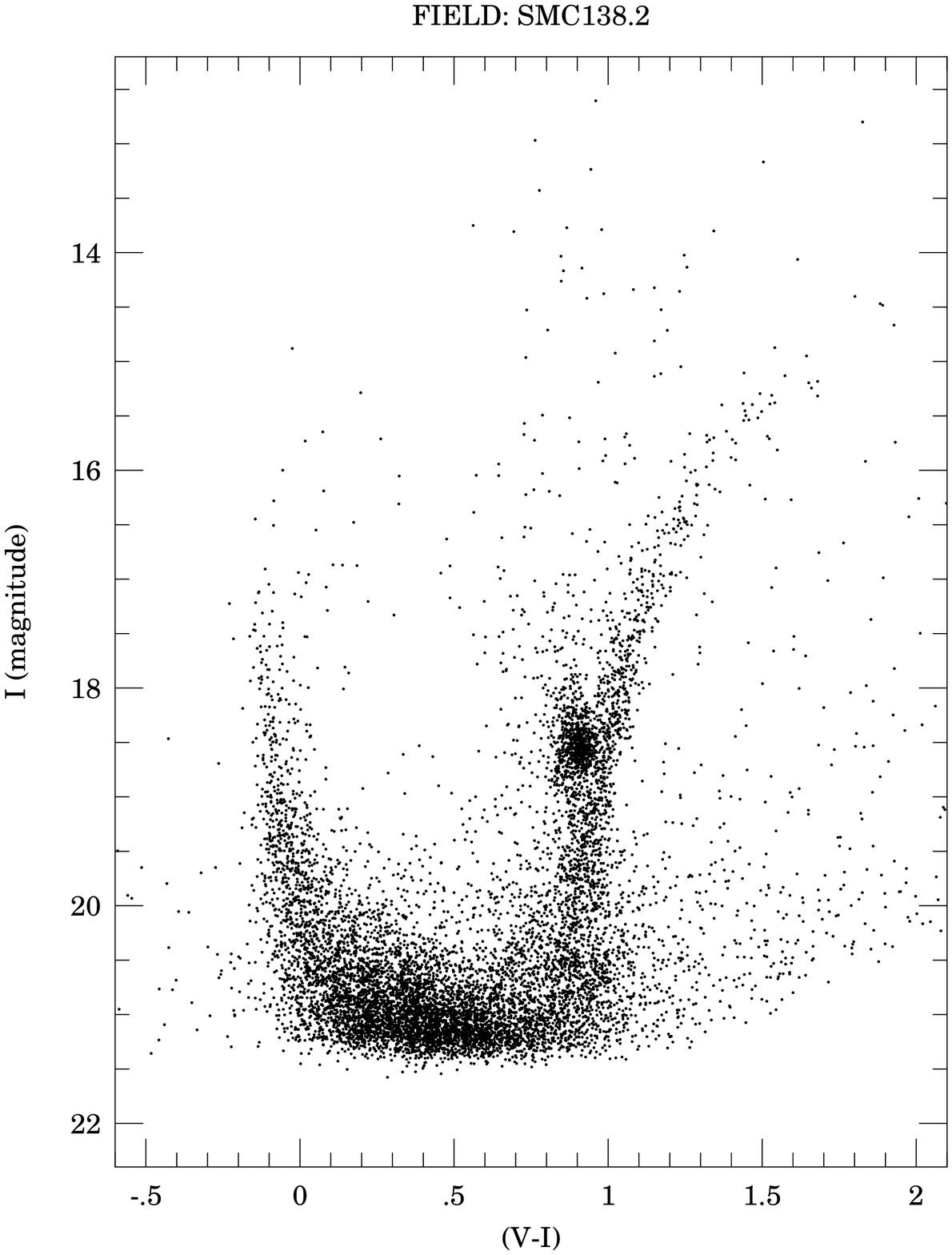}}
\FigCap{Color--magnitude diagram of the western wing subfield SMC138.2}
\end{figure}
\begin{figure}[p]
\centerline{\includegraphics[width=14.5cm, bb=10 50 560 750]{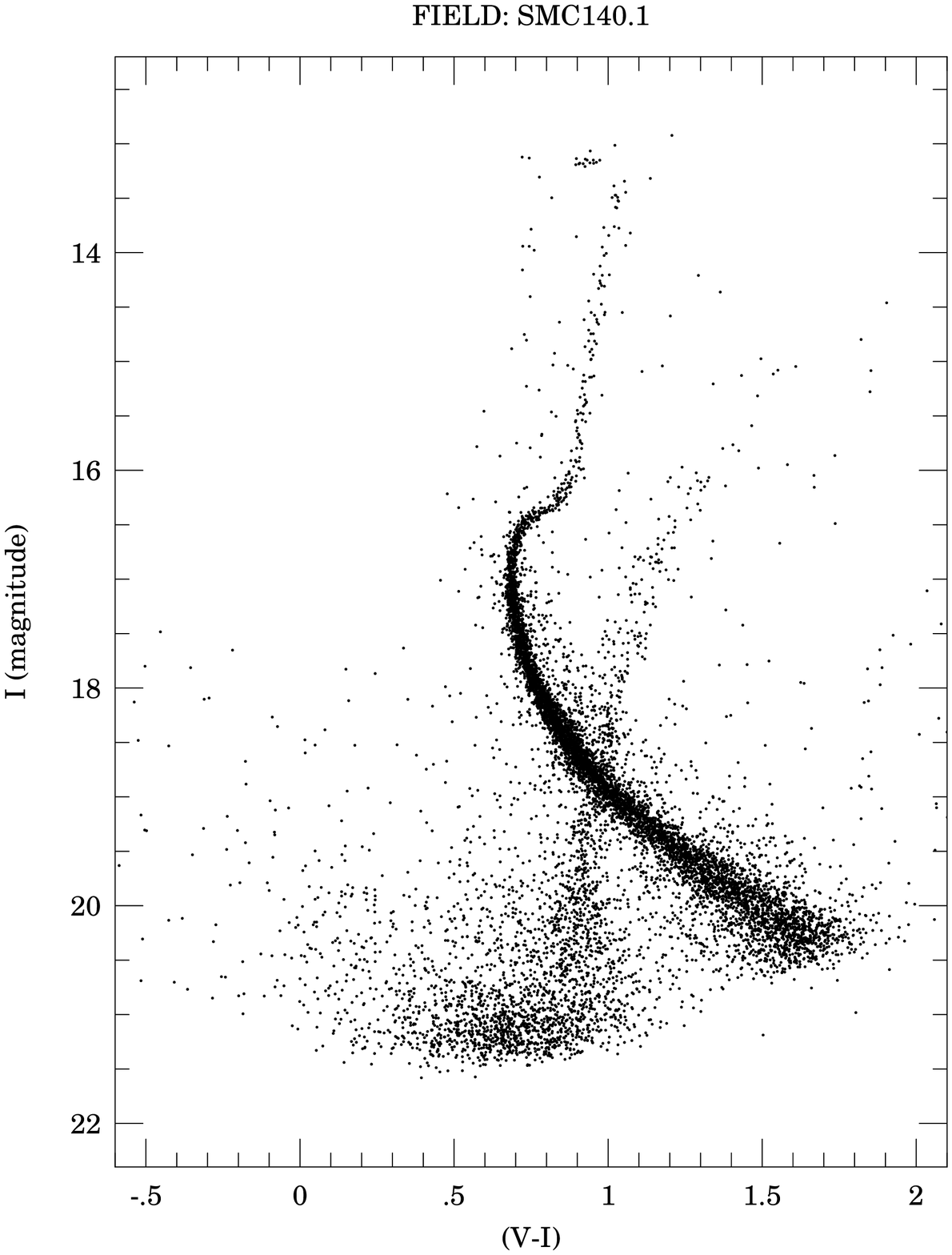}}
\FigCap{Color--magnitude diagram of the field SMC140.1 covering outer
parts of the 47~Tuc globular cluster.}
\end{figure}
\begin{figure}[p]
\centerline{\includegraphics[width=14.5cm, bb=10 50 560 750]{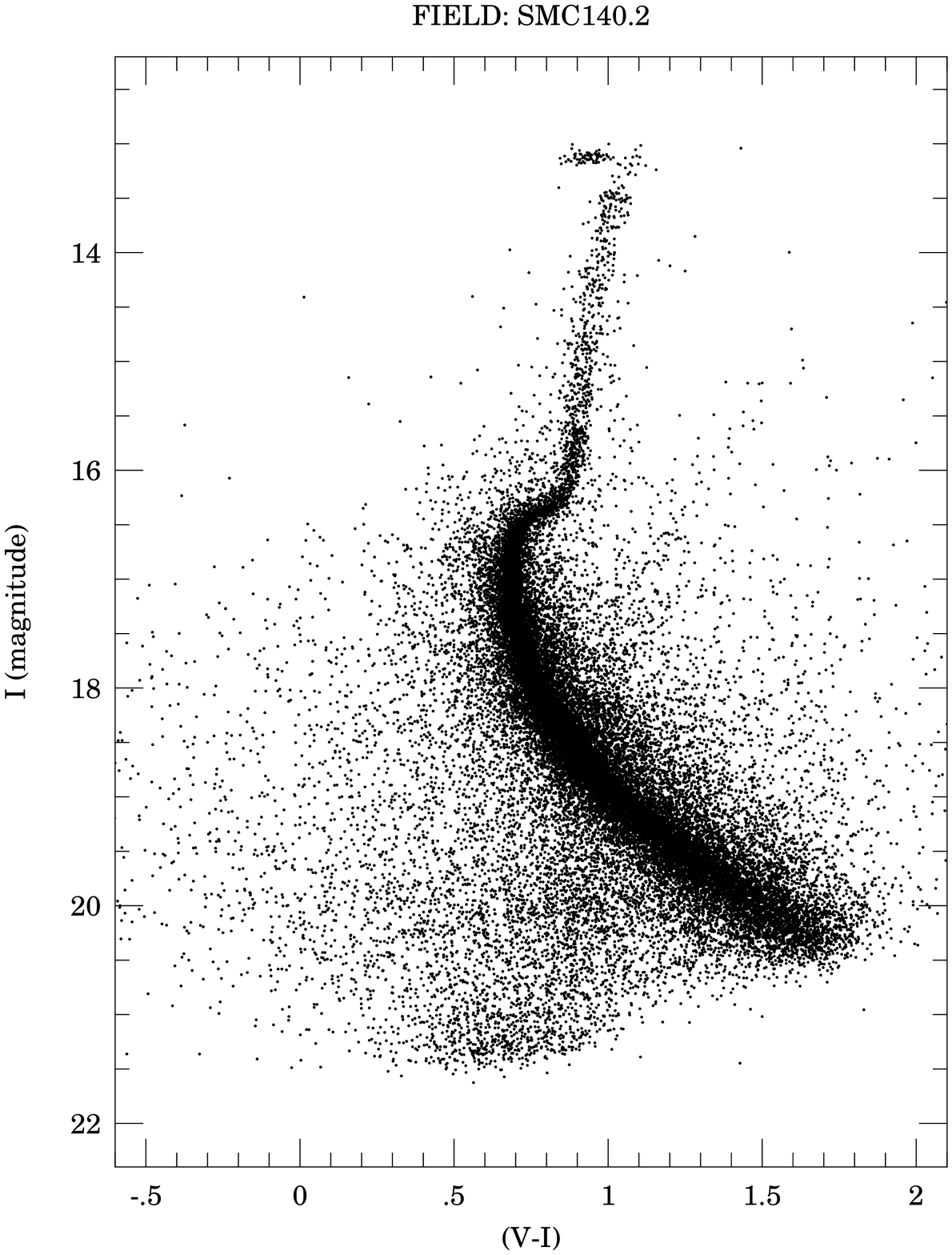}}
\FigCap{Color--magnitude diagram of the field SMC140.2 covering part of
the core of the 47~Tuc globular cluster.}
\end{figure}

\newpage
Figs.~6--8 present, as an example of direct application of the OGLE-III
maps, color--magnitude diagrams (CMDs) constructed for a few selected
subfields from different parts of the SMC. They include the central bar
field, as well as less dense regions of this galaxy from both sides of the
bar. It is believed that the SMC has a considerable depth along the line of
sight. This can indeed be easily noted comparing position of the red clump
in the fields located on the opposite sides of the SMC bar (Figs.~7 and~8).

The SMC140 field is of particular interest. It was centered on the
Galactic globular cluster 47~Tuc and covers practically the whole
cluster. Fig.~9 presents spectacular CMD the outer regions of 47~Tuc
(subfield SMC140.1) superimposed on the CMD of the SMC background stars
while Fig.~10 -- the CMD of the central parts of 47~Tuc (subfield
SMC140.2). Very precise, unique and precisely calibrated photometry of
47 Tuc can be used for many scientific projects related to this cluster.

It should be noted, however, that the standard OGLE-III data pipeline is
not optimized for the extremely dense central regions of the cluster with
high stellar gradients. Therefore the presented photometry in the very
central core regions of 47~Tuc is less complete than could be achieved with
optimized procedures. We plan to rereduce the collected OGLE-III 47~Tuc
images with the optimized software to obtain the photometry as complete as
possible in the core of this cluster.

It is worth noting that OGLE-III maps also contain photometry of many other
stellar clusters located in the SMC.

\section{Data Availability}
The OGLE-III Photometric Maps of the SMC are available to the astronomical
community from the OGLE Internet Archive:

\begin{center}
{\it http://ogle.astrouw.edu.pl}\\
{\it ftp://ftp.astrouw.edu.pl/ogle3/maps/smc/}
\end{center}

Beside tables with photometric data and astrometry for each of the
subfields also the {\it I}-band reference images are included. Usage of the
data is allowed under the proper acknowledgment to the OGLE project.

\Acknow{This paper was partially supported by the following Polish MNiSW
grants: N20303032/4275 to AU and NN203293533 grant to IS and by the
Foundation for Polish Science through the Homing (Powroty) Program.}


\begin{references}
\refitem{Chiosi, E., and Vallenari, A.}{2007}{\AA}{466}{165}
\refitem{Coe, M.J., Edge, W.R.T., Galache, J.L., and McBride, V.A.}{2005}{\MNRAS}{356}{502}
\refitem{Evans, C.J., Howarth, I.D., Irwin, M.J., Burnley, A.W., and
Harries, T.J.}{2004}{\MNRAS}{353}{601}
\refitem{Haberl, F., Eger, P., and Pietsch, W.}{2008}{\AA}{489}{327}
\refitem{Pojma{\'n}ski, G.}{1997}{\Acta}{47}{467}
\refitem{Udalski, A., Szyma\'nski, M., Kubiak, M., Pietrzy\'nski, G., 
Wo\'zniak, P., and {\.Z}ebru\'n, K.}{1998}{\Acta}{48}{147}
\refitem{Udalski, A., Szyma\'nski, M., Kubiak, M., Pietrzy\'nski, G.,
Soszy{\'n}ski, I., Wo\'zniak, P., and {\.Z}ebru\'n,
K.}{2000}{\Acta}{50}{307}
\refitem{Udalski, A., Szyma\'nski, M., Kubiak, M., Pietrzy\'nski, G.,
Soszy{\'n}ski, I., Wo\'zniak, P., {\.Z}ebru\'n, K., Szewczyk, O., and
Wyrzykowski, {\L}.}{2002}{\Acta}{52}{217}
\refitem{Udalski, A.}{2003}{\Acta}{53}{291}
\refitem{Udalski, A., Szyma\'nski, M.K., Soszy\'nski, I., and Poleski.
R.}{2008a}{\Acta}{58}{69}
\refitem{Udalski, A., Soszy{\'n}ski, I., Szyma{\'n}ski, M.K., Kubiak,
M., Pietrzy{\'n}ski, G., Wyrzykowski, {\L}., Szewczyk,~O., Ulaczyk, K.,
and Poleski. R.}{2008b}{\Acta}{58}{89}
\end{references}
\end{document}